# Synchronization of Weak Signals in Dynamic Systems

Mahmut AKILLI[1*]

[1*] Istanbul University, Institute of Science, Department of Physics, Istanbul, Turkey
E-mail: akillimahmut@yahoo.com.tr

**Abstract**

The present study proposes a methodology that combines the 'Duffing oscillator system' and the 'Kuramoto oscillator network' to explore the synchronization of weak signals in dynamic systems. The first step of the procedure is to detect weak periodic or quasi-periodic signals in noisy data collected from the quantifiable processes of any dynamical system using the Duffing oscillator system. The second step is to investigate how the interaction of these weak signals can be synchronized using the Kuramoto oscillator network model. This methodology was applied to seismic signals. The present study has shown that this methodology has great potential for investigating the weak signal synchronisation present within dynamic systems, as evidenced by the analysis of seismic data.

## 1. Introduction

The term 'synchronization' literally means 'sharing common time' or 'occurring at the same time'. The term can be defined as the occurrence of two or more events at the same time. Synchrony can thus be described as the common, regular, and consistent behaviour of interacting elements in a dynamic system over time. Synchronization processes, which have been observed in a variety of natural phenomena, represent a key area of research in a range of fields including physics, technology, chemistry, biology and society. [1-4].

The aim of this study is to present a method to examine the synchronization that may occur as a result of the weak signals' interaction in dynamic systems. The Duffing oscillator system allows us to detect weak signals in noisy data collected from a measured parameter of a dynamical system. The weak signals have a periodic or quasi-periodic nature and are low amplitude. [5-7]. The Kuramoto model is a fundamental mathematical model widely used to explain the synchronization behaviour of multiple interacting oscillators [8]. The weak signal interactions that lead to synchronization states can be studied using the Kuramoto model. Therefore, the present study proposes a hybrid methodology consisting of the Duffing oscillator system and the Kuramoto oscillatory network.

Earthquakes are defined as ground vibrations caused by the sudden release of stress accumulated in the earth's crust as a result of fracturing or sliding movement [9]. The measurements of the ground vibrations, which change over time, are called seismic signals. The proposed methodology will be applied to analyse the potential synchronization of weak signals interacting within seismic signals, as shown in **Figure 1**. In this study, three separate 10-minute seismic data sets were used: the moment the Maras-Pazarcik earthquake occurred in Turkey on 6 February 2023, the 10 minutes preceding it, and the 10 minutes preceding it one month prior [10].

In the first step of the proposed methodology, the weak signal frequencies in the seismic signals were determined using the Duffing oscillator system. Then, the distribution graphs of the angular frequencies of these detected weak seismic signals were plotted. In the second step, it was investigated how the angular frequencies of weak seismic signals can be synchronized using the Kuramoto oscillator network model. In the Kuramoto model, the order parameter graphs of the weak seismic signals were plotted for different values of the coupling parameter, K. The Erdős–Rényi (ER) random network [11], Watts–Strogatz (WS) small world network [12] and Barabási–Albert (BA) scale-free network [13] models were used to obtain these graphs. The results obtained for three seismic data sets show that this method is a valid way to examine the weak signal synchronisation in dynamic systems.

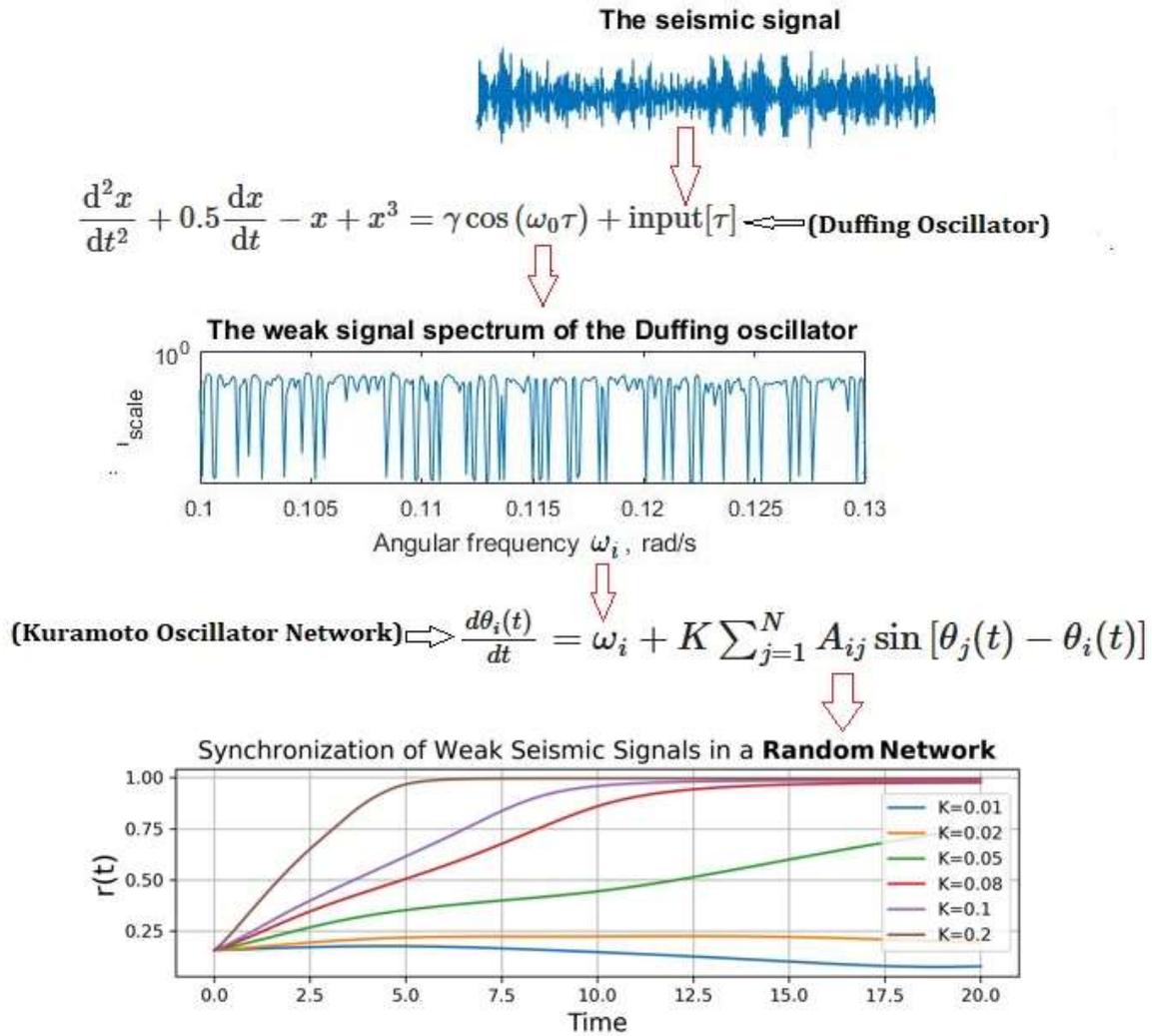

**Figure 1:** Graphic abstract. The methodology combines the 'Duffing oscillator system' and the 'Kuramoto oscillator network' to study the synchronization of weak signals in a dynamic system.

## 2. Methods and Materials
### 2.1. Seismic Data

Seismic data was taken from the KOERI (Bogazici University Kandilli Observatory and Earthquake Monitoring Centre), accessible via http://www.koeri.boun.edu.tr/ [10]. The Mw 7.7 Maras–Pazarcik earthquake of February 6, 2023 occurred at 01:17 UTC, at a depth of 10 km, with an epicentre located at 37.226° N, 37.014° E, ruptured Turkey's East Anatolian Fault Zone (EAFZ) — specifically its Pazarcık segment. The seismic data used in this study were collected from the HHZ (high-gain, high-bandwidth, vertical component) channel at the CEYT (Ceyhan–Adana) seismic station, which is 114 km away from the epicentre of this earthquake. The sampling frequency of the seismic data recording was 100 Hz. **Figure 2** shows the 120-minute seismic signal (a record of the earthquake's vibrations) of the Maras-Pazarcik earthquake.

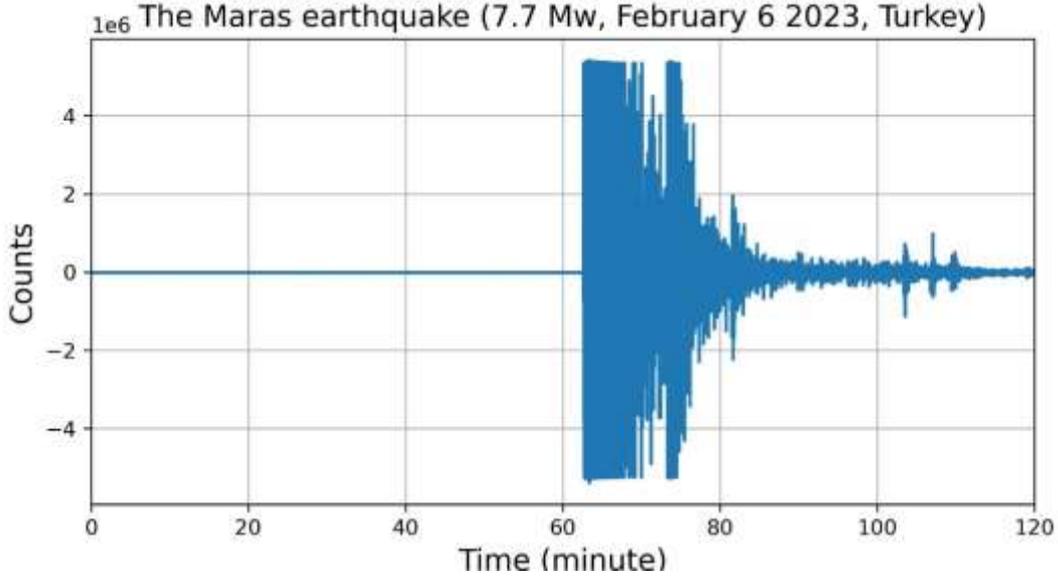

**Figure 2:** The 120-minute seismic signal of the Turkey-Maras (Pazarcik) earthquake, which occurred on 6 February 2023, is presented [10].

### 2.2. Weak Signal Detection in Dynamical Systems

The Duffing oscillator (DO) serves as an optimal mechanism for determining weak periodic or quasi-periodic components within externally applied noisy inputs [6-7]. The input signal added to the DO equation can be expressed as follows;

$$\frac{d^2y}{dt^2} + 0.5\frac{dy}{dt} - y + y^3 = A\cos(t) + \text{input} \qquad (1)$$

For determining the values of weak signal frequencies in the input data, a frequency transformation is applied. Let $t = \omega_0\tau$, here $\omega_0$ denotes the angular frequency. The DO equation (1) is then changed as follows [6];

$$\frac{dy}{d\tau} = \dot{y}(\tau) = \omega_0 z$$

$$\frac{dz}{d\tau} = \dot{z}(\tau) = \omega_0[-0.5z + y - y^3 + \gamma\cos(\omega_0\tau) + input(\tau)] \qquad (2)$$

Here, $A\cos(\omega_0\tau)$ represents the reference signal, and $A$ denotes amplitude. The amplitude value $A_c$, at which the DO system transitions from the threshold of chaotic behaviour to periodic steadiness, is referred to as the bifurcation point. The $\omega_0$ parameter is used to scan for weak signals of unknown frequency within the input data. When the DO system (2) is tuned to the critical chaotic state by adjusting its amplitude value and synchronized such that the weak signal's angular frequency in the externally applied input matches that of the reference signal, the DO transitions into a periodic phase state [6-7].

In order to detect weak signals in a time series collected from any dynamical system, certain adjustments must be made to both the parameters of the DO and the scaling of the time series [5, 14-15]:

  a) Data scaling: The data to be used as input data in the DO equation is rescaled by a coefficient according to the scale limit of the reference signal, $input = c * data$.

b) Adjusting sample frequencies: The DO system and the input data must have the same sampling frequency, $f^{Duffing} = f^{input}$.

c) Adjusting the reference signal's amplitude $A$: The DO must be tuned to the threshold of chaotic using the reference amplitude $A$.

**Example:**

**Figure 3(a)** presents the 10-minute seismic signal of the Turkey-Maras earthquake that occurred on February 6, 2023. The seismic signal was recorded at a sampling frequency of 100 Hz. In order to detect weak signals in this seismic signal;

As illustrated in **Figure 3(b)**, the seismic data was scaled up by multiplying it with a gain factor of 2.5 x 10$^{-10}$.

The sample frequency of the DO is identified by the formula $f^{Duffing} = 100/h$. Here, h is the step interval chosen to solve DO equation (2) numerically. Therefore, if $h$ is set to 0.1, the DO's sample frequency is 1000 Hz. The seismic signal's sampling frequency, $f^{seismic}$, is 100 Hz. To equalize the sampling frequencies, the input signal must be set to 1000 Hz, $f^{Duffing} = f^{input} = 10f^{seismic} = 1000\ Hz$. The weak seismic signal values are then calculated by dividing by 10.

The amplitude parameter $A$ is set to the critical chaotic phase state without an input signal is included in the DO equation (2). For weak signal scanning in the angular frequency range $1.0 \leq \omega_0 \leq 1.1$ rad/s, the amplitude value was fixed to $A = 0.825565$. The relationship between the angular frequency ($\omega_0$) and the frequency (f) in the DO is given by the formula $f = \frac{100\omega_0}{2\pi} = 10f_{seismic}$. Thus, the seismic frequency values correspond to between 1.59 Hz and 1.75 Hz.

**Figure 3(c)** shows the results of weak signal scanning performed by increasing the angular frequency with 0.001 steps. The wavelet scale index [16] measures the degree of aperiodic nature of a signal between 0 and 1. A periodic signal's wavelet scale index is very close to zero; in general, if $i_{scale} <$ 0.002, the signal is periodic [17-20]. As seen in the DO's wavelet scale index spectrum in **Figure 3(c)**, at the angular frequency where the weak signal is found in the seismic signal, the DO transitions to a periodic state, and the scale index value denotes a value close to zero. However, at the angular frequency where there is no weak signal, the DO continues to be chaotic, and the scale index value denotes a number far from zero.

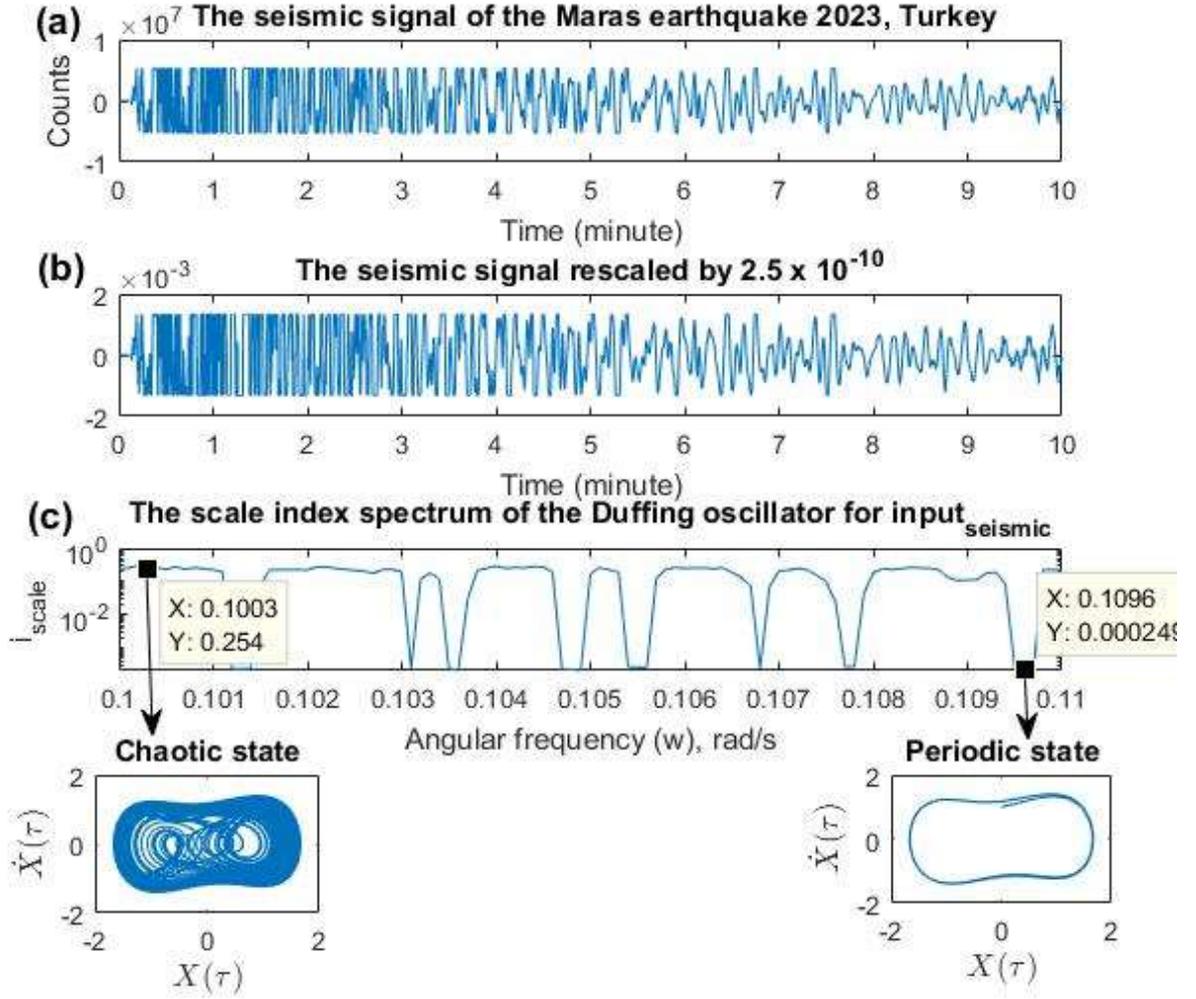

**Figure 3: (a)** The 10-minute seismic signal of the Turkey-Maras (Pazarcik) earthquake that happened on February 6, 2023. [10]. **(b)** The seismic data were rescaled by 2.5 x 10$^{-10}$. **(c)** Wavelet scale index spectrum of the DO: The angular frequencies that correspond to values of the scale index approaching zero indicate weak signals within the seismic signal.

### 2.3. Kuramoto Oscillator Networks

The Kuramoto model [8] describes how coupled oscillators spontaneously synchronize due to interactions. The model presents a mathematical formulation that leads to phase synchronization over time as the interaction result of distributed natural frequencies with different phases through the coupling of a sine function. The mathematical formulation of the Kuramoto model for a network with N oscillators can be written as [21];

$$\frac{d\theta_i}{dt} = \omega_i + K \sum_{j=1}^{N} A_{ij} \sin(\theta_j(t) - \theta_i(t)) \qquad (3)$$

Where $\theta_i$ denotes the oscillator phase $i$ at time $t$. $\omega_i$ is the oscillator's natural frequency $i$, its unit is $radian/time$. N is the number of oscillators. K is the global coupling strength. $A_{ij}$ is the components of adjacency matrix consisting of the numbers 0 and 1 (connectivity between $i$ and $j$, $i \neq j$). If $A_{ij} = 1$, $i$ and $j$ are connected. If $A_{ij} = 0$, there is no connection between them. The phase angles of the *j* and *i*-th oscillators at a given point in time, t, are denoted by $\theta_j(t)$ and $\theta_i(t)$, respectively. Its units are radians. $sin(\theta_j(t) - \theta_i(t))$ measures the phase difference between oscillators $i$ and $j$. Here, small phase differences mean they're almost synchronised; large differences mean they're out of sync. [21].

### 2.3.1. Order Parameter (Synchronization Measure)

The synchronization of the Kuramoto oscillator model is evaluated on a scale from 0 to 1 through the utilisation of the order parameter $r(t)$ [21-22].

$$r(t)e^{i\psi(t)} = \frac{1}{N}\sum_{j=1}^{N} e^{i\theta_j(t)} \qquad (4)$$

$r(t) \in [0,1]$ measures the coherence of the oscillator system. As $r(t)$ approaches zero, the oscillators are incoherent. As $r(t)$ approaches one, they become synchronized. $\psi(t)$ is average phase.

### 2.3.2. Network Topologies

The interaction structure $A_{ij}$ in the model (3) defines the network topology and exerts a significant influence on synchronization. A significant number of network topologies have been developed to explain the synchronisation observed in diverse dynamic systems in nature. In this study, the focus will be on three commonly used network topologies [23]:

*The Erdős–Rényi (ER) random network model:* The ER network [11, 24], as pioneered by Paul Erdös and Alfréd Rényi in 1959, is characterised by a configuration in which every pair of nodes is randomly linked with a fixed probability. The $G(N, p)$ model is generally used. $N$ denotes the total number of nodes in the network model. The symbol $p$ denotes the probability of connection of each edge that can occur between two nodes.

*The Watts–Strogatz (WS) small-world network model*: The WS model [12], which was introduced in 1998 by Duncan Watts and Steven Strogatz, is used to describe small-world networks. The WS model is constructed with three parameters: $N$ indicates the total number of nodes; $k$ indicates the number of nearest neighbours to which each node is connected (in a ring lattice configuration); and $p$, representing the probability of rewiring. To construct a Watts-Strogatz graph, a ring lattice of $N$ nodes with an average degree of $2k$ is first created, and each node is linked to its $k$ closest neighbours. Then, for each edge in the chart, the goal node is reconnected with probability $p$. This gives rise to a perfect ring lattice when $p$ =0, or a random chart when $p$ =1. Network type is small-world if 0< $p$ <1.

*The Barabási–Albert (BA) scale-free network model:* The BA model [13], which was introduced in 1999 by A-L. Barabási and R. Albert, provides a theoretical framework for understanding the evolution of many real-world networks into scale-free structures. In scale-free structures, networks demonstrate a degree distribution which follows a power law instead of the more typical random (Poisson) distribution. The BA model, $G(N, m)$ is constructed with two parameters: $N$ is total nodes, $m$ denotes edges added per new node.

## 3. Results
### 3.1. Weak Signal Detection in Seismic Data

The seismic data, which were recorded before and during the Maras-Pazarcik earthquake in Turkey that occurred on February 6, 2023 [10] were scanned via the DO for detection weak signals. **Figure 4(a)** shows the 10-minute seismic signals recorded one month before the Maras earthquake. **Figure 5(a)** shows the 10-minute seismic signals recorded 10 minutes before the Maras earthquake. **Figure 6(a)** shows the 10-minute seismic signals recorded from the start of the Maras earthquake.

For the detection of weak signals in seismic data, it is necessary to rescale the seismic time series. As demonstrated in **Figures 4(b)** and **5(b)**, the pre-earthquake seismic data were rescaled by multiplying by 2.5 10⁻⁷. As demonstrated in **Figure 6(b)**, the seismic data during the Maras earthquake were rescaled by multiplying by 2.5 10⁻⁹.

To solve the DO equation numerically, the h step interval was set to 0.1. Accordingly, the DO's sample frequency $f^{Duffing}$ is 1000 Hz (100/0.1). The input signal's sample frequency $f^{input}$ must also be set to 1000 Hz, $f^{Duffing} = f^{input} = 10f^{seismic} = 1000\ Hz$. The seismic signal's sample frequency $f^{seismic}$ is 100 Hz. Then, the frequency values of the weak signals detected in the input seismic signal are divided by 10 to calculate the weak seismic signals' frequency values.

The DO system (2) is set to the threshold of chaotic phase using the reference amplitude $A$. In order to keep the oscillator at this threshold of critical chaotic value, the amplitude value $A$ must be subjected to small changes after certain angular frequency intervals [25]. The input signal was checked in 0.001-step increments within the angular frequency range $0.508 \leq \omega_0 \leq 3.2$, and weak signals were found. Here, the seismic frequency values correspond to between 0.808 Hz and 5.093 Hz, the range of seismic angular frequencies is also $0.0508 \leq \omega_0 \leq 0.32$. This is calculated using the formula $f = \frac{100\omega_0}{2\pi} = 10f_{seismic}$.

The scale index spectrums of the DO for the input seismic signals can be observed in **Figures 4(c)**, **5(c)** and **6(c)**. In computation of scale index values, it was used the Haar wavelet with scales between $s_0 = 1$ and $s_1 = 1024$. The scale index values that hang towards zero, corresponding to angular frequencies in the graphs, indicate weak signals detected in the seismic data. **Figures 4(d)**, **5(d)** and **6(d)** show the distribution of weak signals detected in the seismic signals. These figures show the histogram of weak signals and their corresponding Kernel Density Estimation (KDE) [26-27] and Normal Probability Density Distribution (PDF).

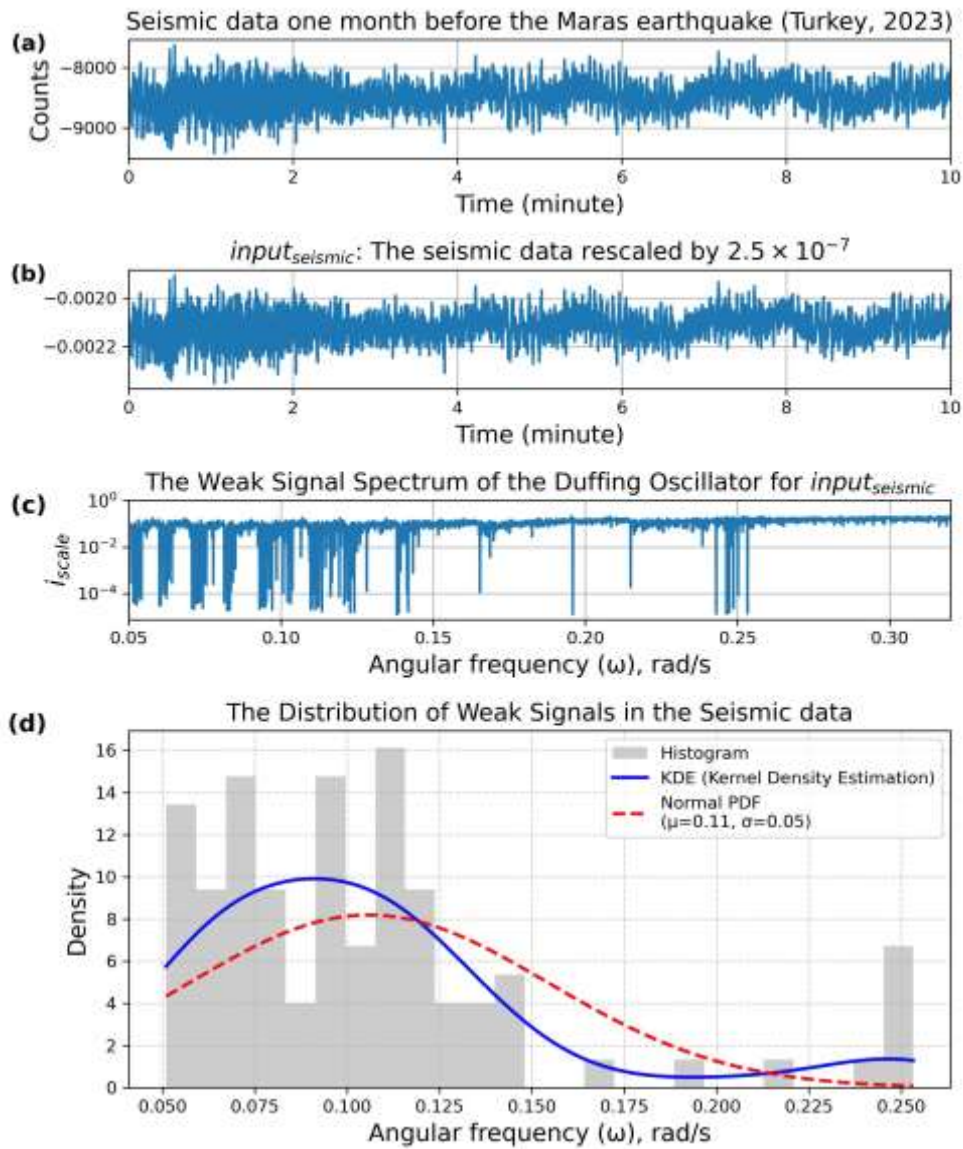

**Figure 4:** (**a**) The 10-minute seismic signal from **one month** before the Maras - Pazarcik earthquake in Turkey that occurred on February 6, 2023 [10]. (**b**) The seismic data were rescaled by 2.5 x $10^{-7}$. (**c**) The DO's wavelet scale index spectrum: The angular frequencies concomitant with the wavelet scale index values which fall to zero are indicative of weak signals within the seismic data. The number of weak seismic signals detected in these data is 90. (**d**) The density distribution of weak seismic signals.

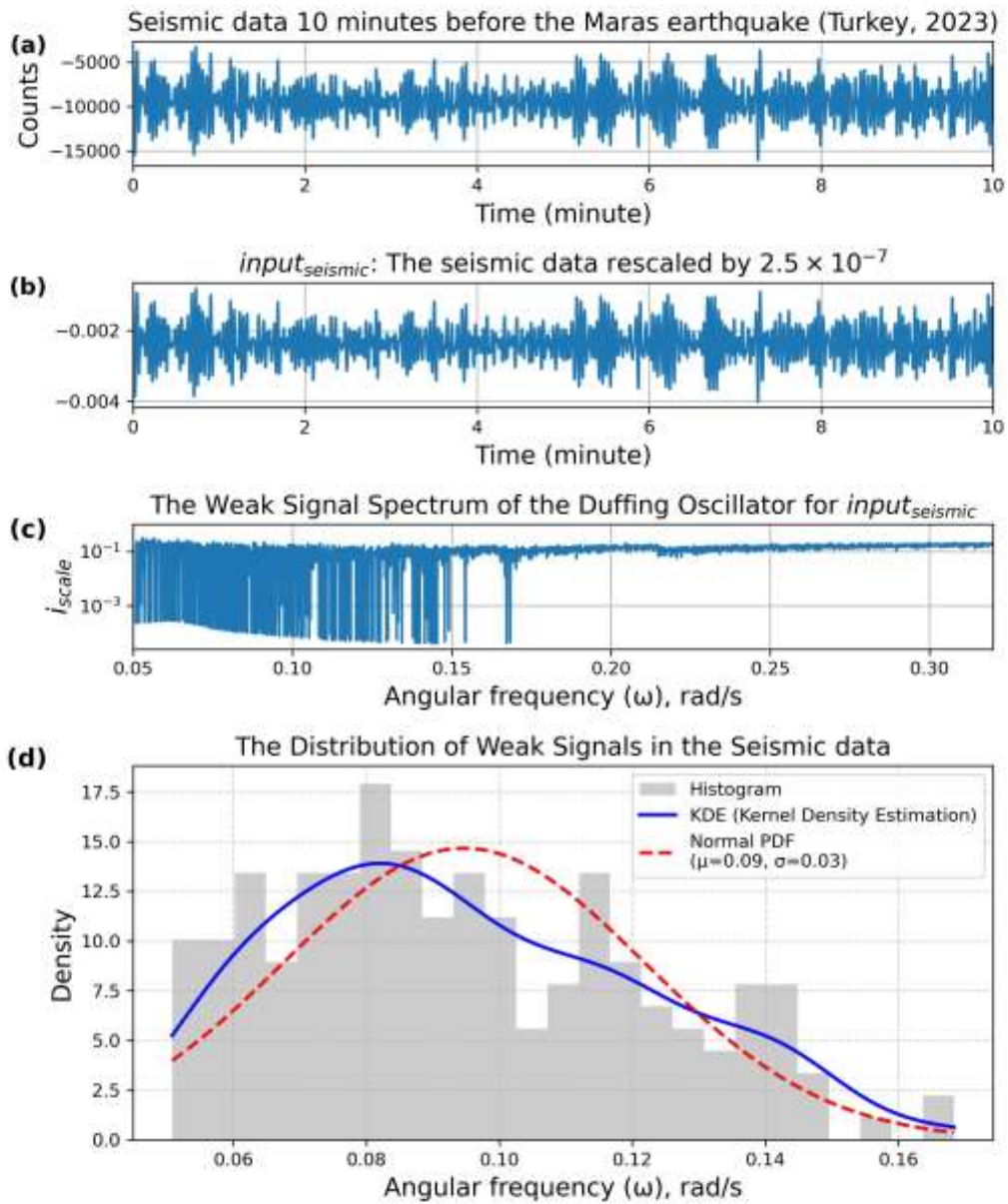

**Figure 5:** (**a**) The 10-minute seismic signal from **10 minutes before** the Maras - Pazarcik earthquake in Turkey that occurred on February 6, 2023 [10]. (**b**) The seismic data were rescaled by $2.5 \times 10^{-7}$. (**c**) The DO's wavelet scale index spectrum: The angular frequencies concomitant with the wavelet scale index values which fall to zero are indicative of weak signals within the seismic data. The number of weak seismic signals detected in these data is 190. (**d**) The density distribution of weak seismic signals.

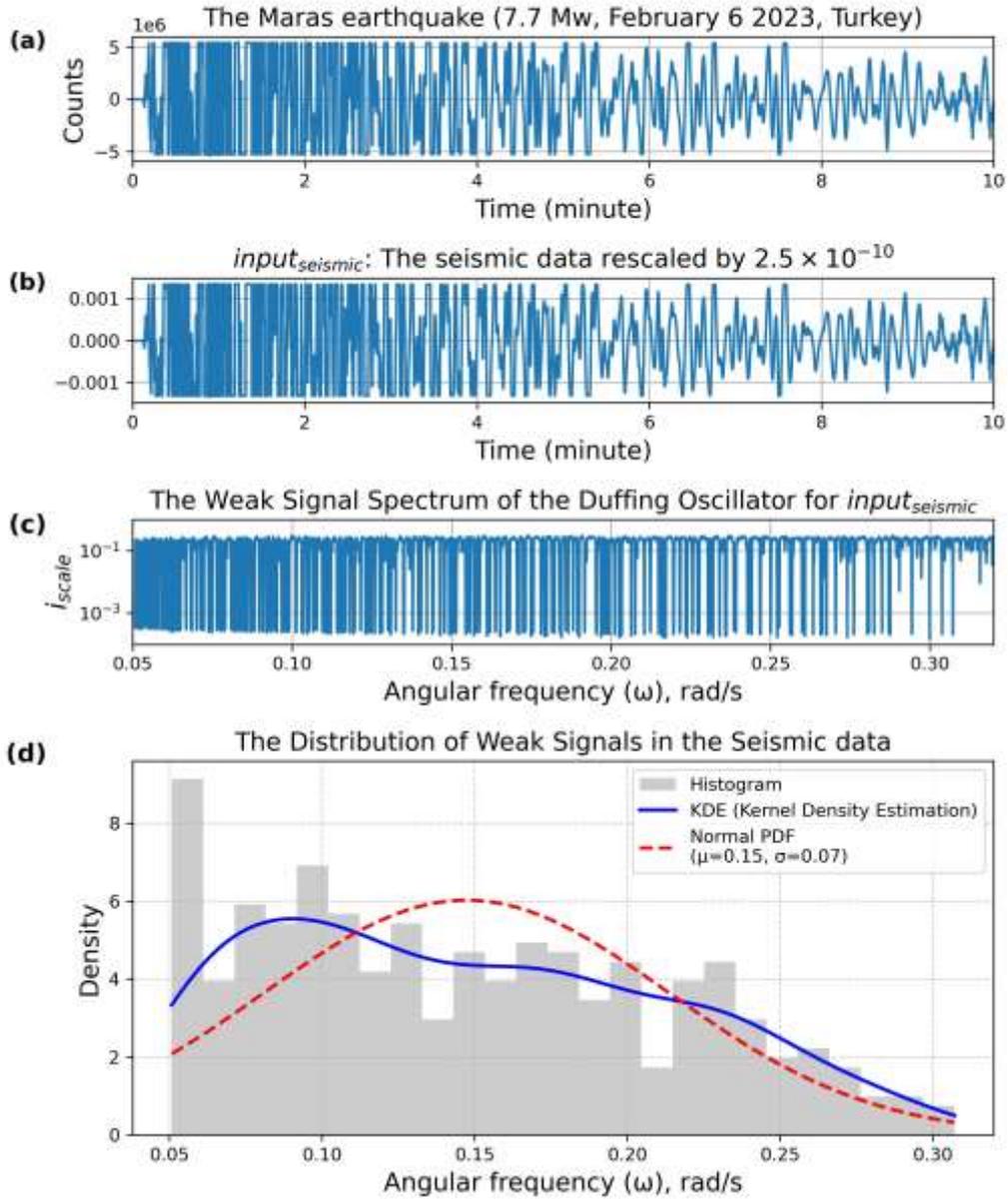

**Figure 6:** (**a**) The 10-minute seismic signal recorded from the moment of the Maras-Pazarcik earthquake that occurred in Turkey on 6 February 2023 [10]. (**b**) The seismic data were rescaled by 2.5 x 10$^{-10}$. (**c**) The DO's wavelet scale index spectrum: The angular frequencies concomitant with the wavelet scale index values which fall to zero are indicative of weak signals within the seismic data. The number of weak seismic signals detected in these data is 395. (**d**) The density distribution of weak seismic signals.

### 3.2. Synchronization of weak seismic signals

In the previous section, the DO system (2) was used to identify the angular frequencies of weak signals in seismic data. In this section, Kuramoto oscillator networks were used to study the formation of synchronization of angular frequencies of weak seismic signals. The angular frequency values of weak seismic signals $\omega_i$ were used instead of the natural frequencies in the Kuramoto model equation (3). The weak signal number detected in the seismic data was used instead of $N$. The positions of the angular frequencies of the weak signals, which were arranged from smallest to largest, were randomly mixed, and negative values were assigned to 30% of them. The phase angles of the weak seismic signals, $\theta_j(t)$ and $\theta_i(t)$, were also randomly generated. The ER random network, the WS small-world network and the BA scale-free network models were used in the Kuramoto equation (3). For these network models, the order parameters $r(t)$ (4), which are measures of synchronization, were calculated for different K coupling values and their graphs were drawn.

**Figure 7(b)** shows the density distribution of the weak signals' angular frequencies detected in the 10-minute seismic data, which were recorded one month prior to the Maras - Pazarcik earthquake in Turkey that occurred on February 6, 2023, (see **Figure 7(a)**). The number of weak seismic signals that can be detected in this data is N=92. Three network models were used to synchronize the angular frequencies of these weak seismic signals $\omega_i$. **Figure 7(c)** presents the order parameters calculated for different K coupling values in the Kuramoto model using the ER random network, $G(N, p) = G(92, 0.1)$. **Figure 7(d)** presents the order parameters calculated for different K coupling values in the Kuramoto model using the WS small-world network, $G(N, k, p) = G(92, 8, 0.3)$. **Figure 7(e)** presents the order parameters calculated for different K coupling values in the Kuramoto model using the BA scale-free network, $G(N, m) = G(92, 4)$.

**Figure 8(b)** shows the density distribution of the weak signals' angular frequencies detected in the 10-minute seismic data, which were recorded 10 minutes prior to the Maras - Pazarcik earthquake (see **Figure 8(a)**). The number of weak seismic signals that can be detected in this data is N=190. Three network models were used to synchronize the angular frequencies of these weak seismic signals $\omega_i$. **Figure 8(c)** presents the order parameters calculated for different K coupling values in the Kuramoto model using the ER random network, $G(N, p) = G(190, 0.1)$. **Figure 8(d)** presents the order parameters calculated for different K coupling values in the Kuramoto model using the WS small-world network, $G(N, k, p) = G(190, 8, 0.3)$. **Figure 8(e)** presents the order parameters calculated for different K coupling values in the Kuramoto model using the BA scale-free network, $G(N, m) = G(190, 4)$.

**Figure 9(b)** shows the density distribution of the weak signals' angular frequencies detected in the 10-minute seismic data recorded the moment when the Maras - Pazarcik earthquake occurred, (see **Figure 9(a)**). The number of weak seismic signals that can be detected in this data is N=395. Three network models were used to synchronize the angular frequencies of these weak seismic signals $\omega_i$. **Figure 9(c)** presents the order parameters calculated for different K coupling values in the Kuramoto model using the ER random network, $G(N, p) = G(395, 0.1)$. **Figure 9(d)** presents the order parameters calculated for different K coupling values in the Kuramoto model using the WS small-world network, $G(N, k, p) = G(395, 8, 0.3)$. **Figure 9(e)** presents the order parameters calculated for different K coupling values in the Kuramoto model using the BA scale-free network, $G(N, m) = G(395, 4)$.

**Table 1** presents the critical coupling values K, which are calculated in various network models, in order to synchronise the weak signals detected in the seismic signals recorded at different times, as can be seen in **Figures 7**, **8** and **9**. Here, the critical coupling value K is the first value that brings the order parameter close to 1, $r(t) \to 1$.

**Table 1**: Calculated critical coupling values K in Kuramoto oscillator networks to synchronize the angular frequencies of weak seismic signals for the 2023 Maras-Pazarcik earthquake.

| Network models | One month ago | 10 minutes ago | The moment of the earthquake |
|---|---|---|---|
| The ER random network | 0.08 | 0.05 | 0.02 |
| The WS small-world network | 0.15 | 0.15 | 0.15 |
| The BA scale-free network | 0.15 | 0.1 | 0.1 |

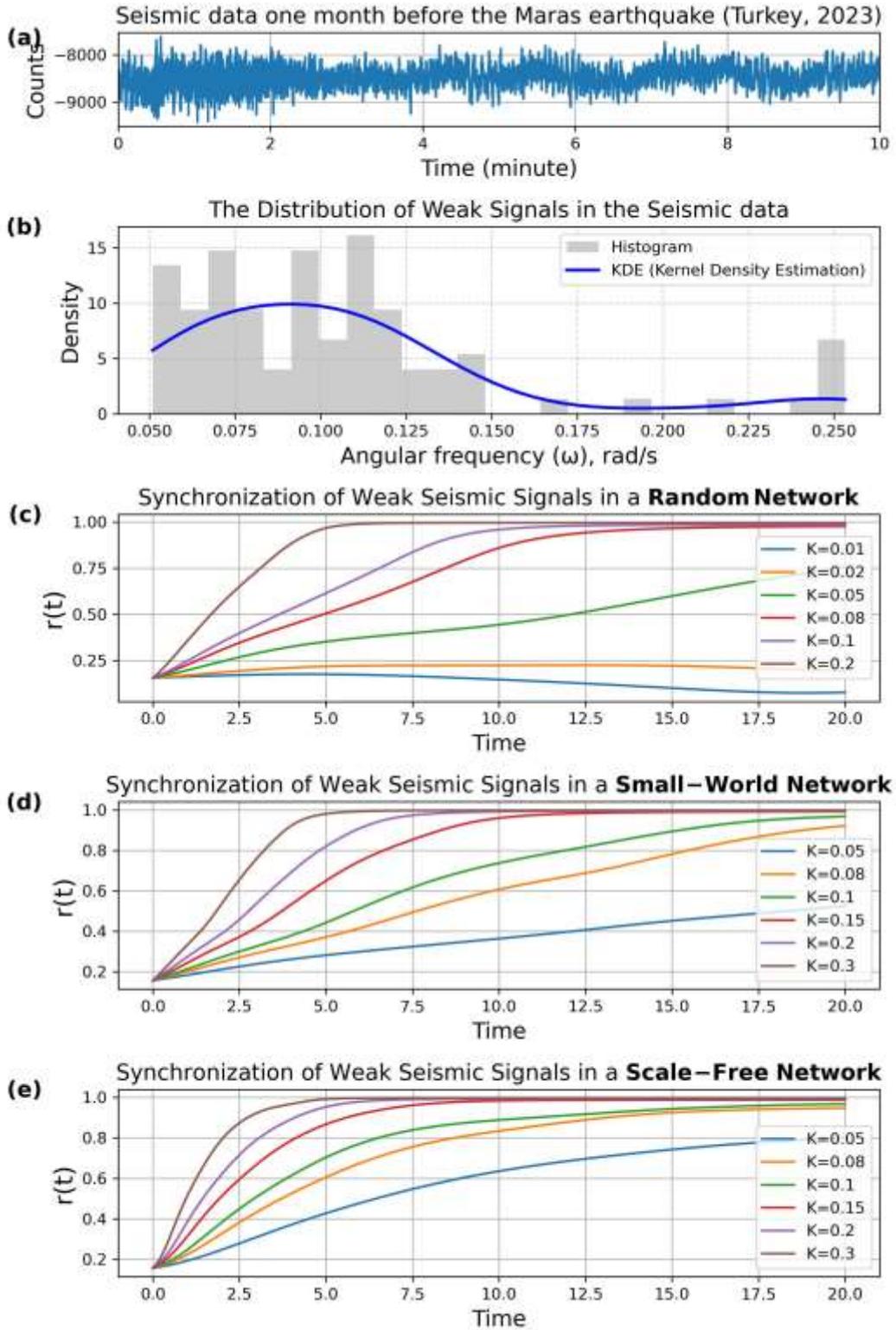

**Figure 7:** (**a**) The 10-minute seismic signal from **one month** before the Maras - Pazarcik earthquake in Turkey that occurred on February 6, 2023 [10]. (**b**) The density distribution of weak signals detected in these seismic data, N=92. (**c**) The order parameters of the weak seismic signals calculated for different K coupling values in the Kuramoto model using **the ER random network**, $G(N, p) = G(92, 0.1)$. (**d**) The order parameters of the weak seismic signals calculated for different K coupling values in the Kuramoto model using **the WS small-world network**, $G(N, k, p) = G(92, 8, 0.3)$. (**e**) The order parameters of the weak seismic signals calculated for different K coupling values in the Kuramoto model using **the BA scale-free network**, $G(N, m) = G(92, 4)$.

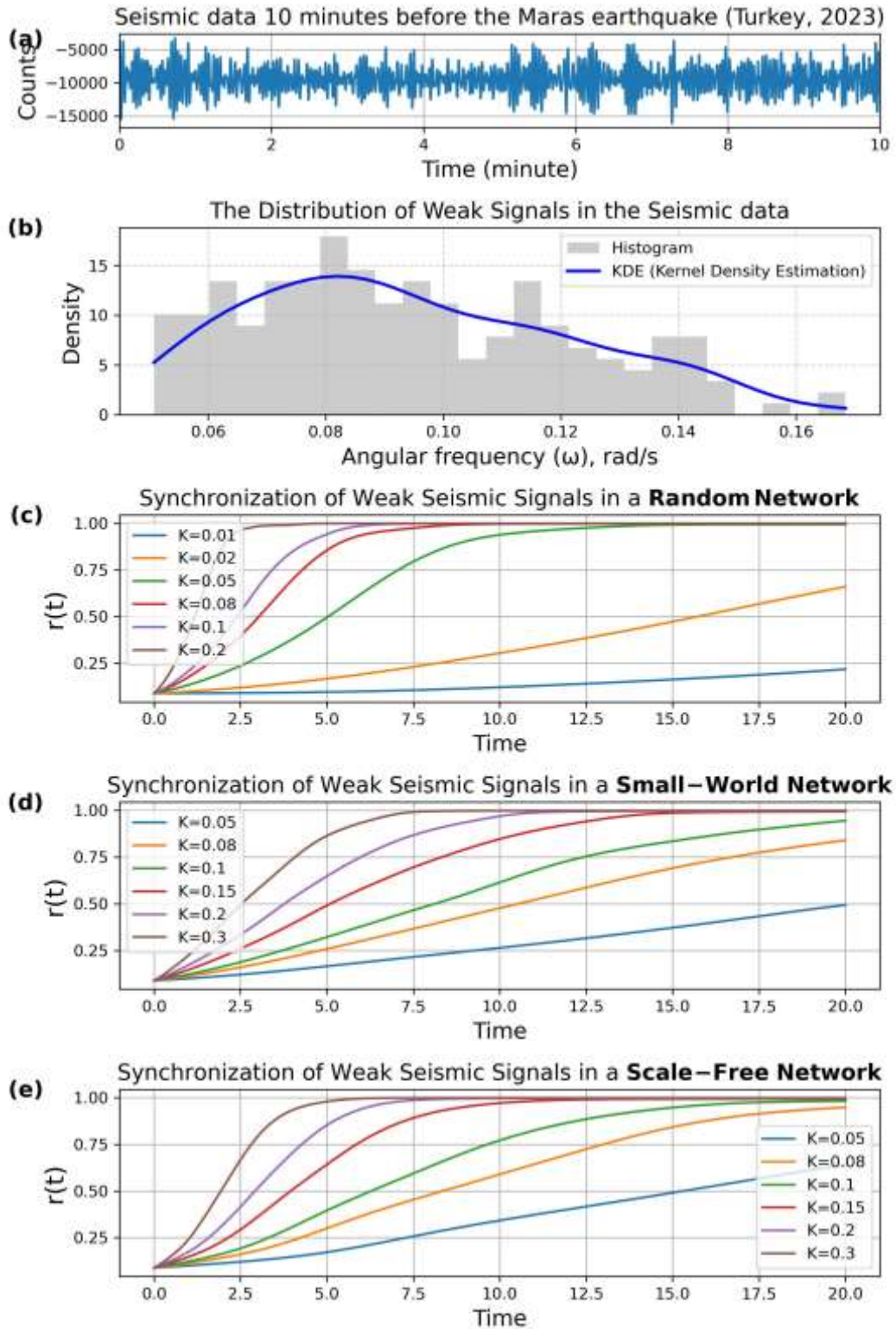

**Figure 8:** (**a**) The 10-minute seismic signal from **10 minutes** before the Maras - Pazarcik earthquake in Turkey that occurred on February 6, 2023 [10]. (**b**) The density distribution of weak signals detected in these seismic data, N=190. (**c**) The order parameters of the weak seismic signals calculated for different K coupling values in the Kuramoto model using **the ER random network**, $G(N, p) = G(190, 0.1)$. (**d**) The order parameters of the weak seismic signals calculated for different K coupling values in the Kuramoto model using **the WS small-world network**, $G(N, k, p) = G(190, 8, 0.3)$. (**e**) The order parameters of the weak seismic signals calculated for different K coupling values in the Kuramoto model using **the BA scale-free network**, $G(N, m) = G(190, 4)$.

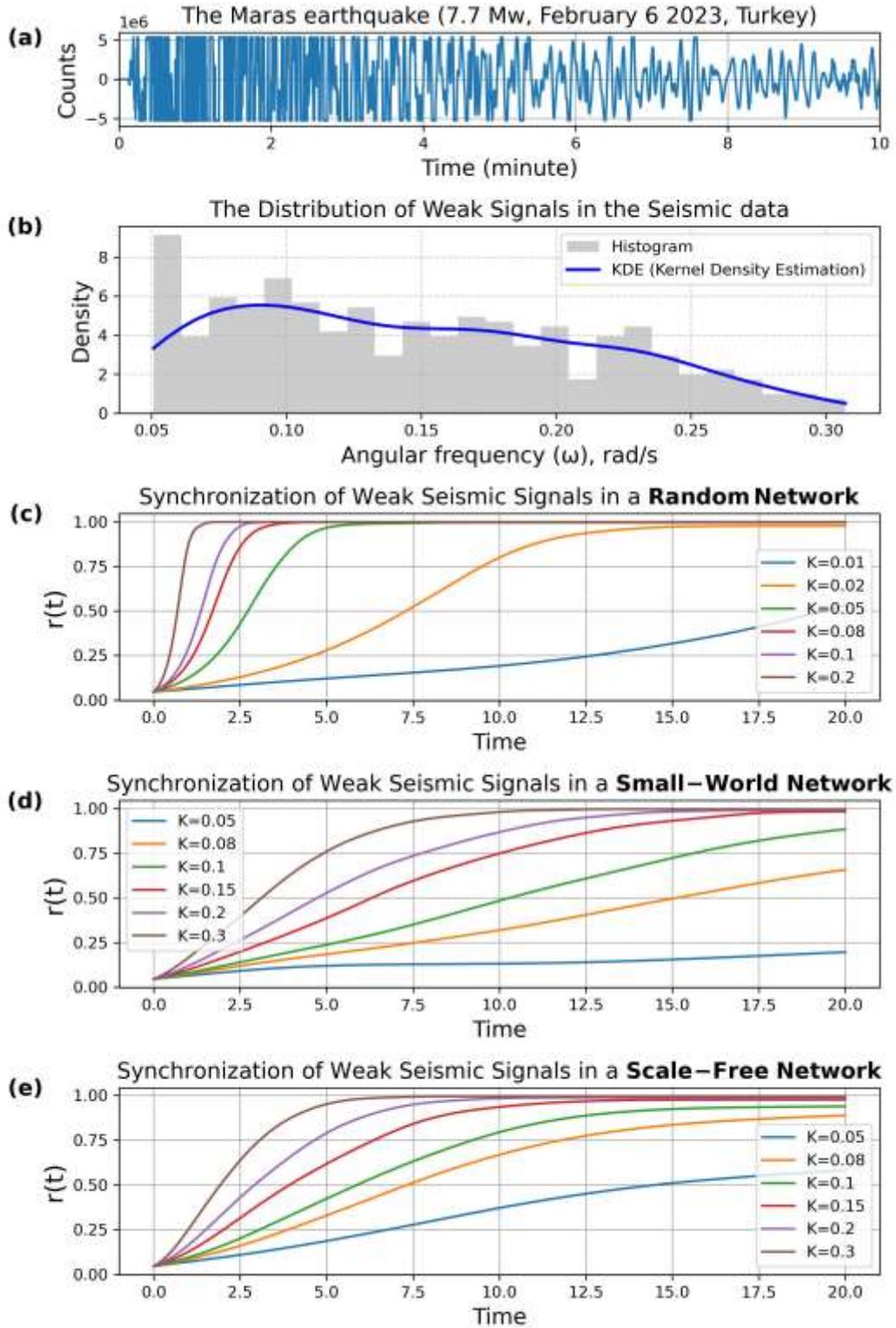

**Figure 9:** (**a**) The 10-minute seismic signal recorded from the **moment** of the Maras-Pazarcik earthquake that occurred in Turkey on 6 February 2023 [10]. (**b**) The density distribution of weak signals detected in these seismic data, N=395. (**c**) The order parameters of the weak seismic signals calculated for different K coupling values in the Kuramoto model using **the ER random network**, $G(N, p) = G(395, 0.1)$. (**d**) The order parameters of the weak seismic signals calculated for different K coupling values in the Kuramoto model using **the WS small-world network**, $G(N, k, p) = G(395, 8, 0.3)$. (**e**) The order parameters of the weak seismic signals calculated for different K coupling values in the Kuramoto model using **the BA scale-free network**, $G(N, m) = G(395, 4)$.

## 4. Discussion

The objective of this study is to study the synchronisation of weak signals in dynamic systems through the utilisation of a methodology that integrates the 'DO system' and the 'Kuramoto oscillator network'. The first step is to detect weak signals in noisy data collected from a measured parameter of a dynamical system, using the DO system. The second step is to examine the synchronization of these weak signals' angular frequencies through the Kuramoto oscillator network. Seismic signals were used for this application.

Three separate 10-minute seismic signals were examined: At the time of the 2023 Maras-Pazarcik earthquake occurred, 10 minutes before it, and one month before it. **Figures 4**, **5** and **6** show the weak signals detected in these seismic signals, which were scanned in the range from 0.0508 rad/s to 0.32 rad/s via the DO system. As illustrated in **Figures 4(d)**, **5(d)** and **6(d)**, the distributions of the weak signals found in these three separate seismic signals are different from one another.

**Figures 7**, **8**, and **9** show the order parameters of the weak seismic signals calculated for different values of $K$ coupling in the Kuramoto oscillator network model using the BA scale-free network, the WS small-world network and the ER random network. **Table 1** demonstrates the critical values of $K$ that bring firstly the order parameters to 1 for the synchronization of weak seismic signals at three different time periods: one month before the earthquake, 10 minutes before the earthquake, and at the time of the earthquake, respectively.

As is known, as the moment of the earthquake approaches, the synchronisation of ground vibrations is expected to increase [28-32]. Therefore, the coupling value $K$ calculated for 10 minutes before the earthquake is expected to be smaller than the coupling value calculated one month before the earthquake. According to this approach, the ER random and the BA scale-free network models are compatible with these two situations. However, there is no change in the critical coupling value $K$ of the WS small-world network model. Looking at the results overall, it seems that the ER random network model is better at synchronising weak seismic signals.

The hybrid methodology, consisting of the Duffing oscillator system and the Kuramoto oscillator network model, can be quite useful for investigating weak signal synchronisation in dynamic systems, as seen in its application to seismic data. This methodology can play a particularly important role in investigating synchrony in the brain by applying it to the EEG signal.

**Data Availability Statement** The data will be made available upon reasonable request. Seismic data were obtained from the Kandilli Observatory and Earthquake Research Institute (KOERI), which provides open access to its data through its official website. KOERI states on its website that its seismic data may be used for scientific research without requiring formal permission from the institution. [Online]. Available: https://www.koeri.boun.edu.tr .

**References**

[1] Pecora, Louis M., and Thomas L. Carroll. "Master stability functions for synchronized coupled systems." Physical review letters 80.10 (1998): 2109.

[2] Pikovsky, Arkady, et al. "Synchronization: A universal concept in nonlinear sciences." Cambridge University Press, 2001.

[3] Strogatz, Steven. "Sync: The emerging science of spontaneous order." (2004).

[4] Pecora, Louis M., and Thomas L. Carroll. "Synchronization of chaotic systems." Chaos: An Interdisciplinary Journal of Nonlinear Science 25.9 (2015).

[5] Akilli, Mahmut, Nazmi Yilmaz, and Kamil Gediz Akdeniz. "Automated system for weak periodic signal detection based on Duffing oscillator." IET Signal Processing 14.10 (2020): 710-716.

[6] Wang, Guanyu, et al. "The application of chaotic oscillators to weak signal detection." IEEE Transactions on industrial electronics 46.2 (1999): 440-444.

[7] Wang, Guanyu, and Sailing He. "A quantitative study on detection and estimation of weak signals by using chaotic Duffing oscillators." IEEE Transactions on Circuits and Systems I: Fundamental Theory and Applications 50.7 (2003): 945-953.

[8] Kuramoto, Yoshiki. "Self-entrainment of a population of coupled non-linear oscillators." International symposium on mathematical problems in theoretical physics: January 23–29, 1975, kyoto university, kyoto/Japan. Berlin, Heidelberg: Springer Berlin Heidelberg, 2005.

[9] Stein, Seth, and Michael Wysession. An introduction to seismology, earthquakes, and earth structure. John Wiley & Sons, 2009.

[10] Kandilli Observatory and Earthquake Research Institute (KOERI), "Kahramanmaraş (Pazarcık) earthquake: Moment magnitude and aftershock activity data," Boğaziçi University, Feb. 6, 2023. [Online]. Available: https://www.koeri.boun.edu.tr .

[11] Erdős, P., and A. Rényi. "On Random Graphs I." Publicationes Mathematicae Debrecen, vol. 6 (1959), pp. 290–297.

[12] Watts, Duncan J., and Steven H. Strogatz. "Collective dynamics of 'small-world'networks." nature 393.6684 (1998): 440-442.

[13] Barabási, Albert-László, and Réka Albert. "Emergence of scaling in random networks." science 286.5439 (1999): 509-512.

[14] Akilli, Mahmut. "Detecting weak periodic signals in EEG time series." Chinese Journal of Physics 54.1 (2016): 77-85.

[15] Akilli, Mahmut, and Nazmi Yilmaz. "Study of weak periodic signals in the EEG signals and their relationship with postsynaptic potentials." IEEE Transactions on Neural Systems and Rehabilitation Engineering 26.10 (2018): 1918-1925.

[16] Benítez, Rafael, Vicente J. Bolós, and M. E. Ramírez. "A wavelet-based tool for studying non-periodicity." Computers & Mathematics with Applications 60.3 (2010): 634-641.

[17] Akıllı, Mahmut, and Nazmi Yılmaz. "Windowed scalogram entropy: wavelet-based tool to analyze the temporal change of entropy of a time series." The European Physical Journal plus 136.11 (2021): 1165.

[18] Yılmaz, Nazmi, et al. "Application of the nonlinear methods in pneumocardiogram signals." Journal of Biological Physics 46.2 (2020): 209-222.

[19] Akıllı, Mahmut, Nazmi Yılmaz, and K. Gediz Akdeniz. "Study of the q-Gaussian distribution with the scale index and calculating entropy by normalized inner scalogram." Physics Letters A 383.11 (2019): 1099-1104.

[20] Akıllı, Mahmut, Nazmi Yılmaz, and K. Gediz Akdeniz. "The 'wavelet'entropic index q of non-extensive statistical mechanics and superstatistics." Chaos, Solitons & Fractals 150 (2021): 111094.

[21] Acebrón, Juan A., et al. "The Kuramoto model: A simple paradigm for synchronization phenomena." Reviews of modern physics 77.1 (2005): 137-185.

[22] Brede, Markus. "Synchrony-optimized networks of non-identical Kuramoto oscillators." Physics Letters A 372.15 (2008): 2618-2622.

[23] Rodrigues, Francisco A., et al. "The Kuramoto model in complex networks." Physics Reports 610 (2016): 1-98.

[24] Erdős, P., and A. Rényi. "The Evolution of Random Graphs." Magyar Tudományos Akadémia Matematikai Kutató Intézet Közleményei, vol. 5 (1960), pp. 17–61.

[25] Yilmaz, Nazmi, and Mahmut Akilli. "Weak periodic signal identification in low amplitude seismic waves based on chaotic oscillator." Journal of Physics: Complexity (2025).

[26] Silverman, Bernard W. Density estimation for statistics and data analysis. Routledge, 2018.

[27] Akilli, Mahmut, and Nazmi Yilmaz. "EEG Brain mapping based on the Duffing oscillator." arXiv preprint arXiv:2509.00618 (2025).

[28] Ohtani, Makiko, Nobuki Kame, and Masao Nakatani. "Synchronization of megathrust earthquakes to periodic slow slip events in a single-degree-of-freedom spring-slider model." Scientific Reports 9.1 (2019): 8285.

[29] Karabulut, Hayrullah, Michel Bouchon, and Jean Schmittbuhl. "Synchronization of small-scale seismic clusters reveals large-scale plate deformation." Earth, Planets and Space 74.1 (2022): 1-11.

[30] Zhang, Huai, et al. "Correlation of earthquake occurrence among major fault zones in the eastern margin of the Tibetan Plateau through Big Data Analysis." The Innovation Geoscience (2025): 100145-1.

[31] Goldfinger, C., et al. "Unravelling the dance of earthquakes: Evidence of partial synchronization of the northern San Andreas fault and Cascadia megathrust." Geosphere 21.6 (2025): 1132-1180.

[32] Vasudevan, Kris, Michael Cavers, and Antony Ware. "Earthquake sequencing: Chimera states with Kuramoto model dynamics on directed graphs." Nonlinear Processes in Geophysics 22.5 (2015): 499-512.